\documentclass{article}
\usepackage{ijcai22}

\usepackage{times}
\usepackage{soul}
\usepackage{url}
\usepackage{hyperref}
\usepackage[utf8]{inputenc}
\usepackage[small]{caption}
\usepackage{graphicx}
\usepackage{amsmath}
\usepackage{amsthm}
\usepackage{booktabs}
\usepackage{algorithm}
\usepackage{algorithmic}
\usepackage{mathrsfs}
\usepackage{textpos}

\usepackage{amsmath,amssymb,amsfonts}
\usepackage{algorithm}
\usepackage{algorithmic}
\usepackage{graphicx}
\usepackage{textcomp}
\usepackage{xcolor}
\usepackage{subfigure}
\usepackage{multirow}
\usepackage{bigstrut}

\newcommand{\liu}[1]{{\color{black}{#1}}}

\newcommand{\TheName}{FedDUAP}
\newcommand{\TheAlgoName}{FedDU}
\newcommand{\ThePruneName}{FedAP}

\title{\TheName{}: \underline{Fed}erated Learning with \underline{D}ynamic \underline{U}pdate and \underline{A}daptive \underline{P}runing Using Shared Data on the Server}

\author{
Hong Zhang$^{1}$\setcounter{footnote}{1}\footnote{H. Zhang and J. Liu contributed equally to the paper. This work was done when H. Zhang was an intern at Baidu Inc.}, 
Ji Liu$^{2}$\footnotemark[2]\setcounter{footnote}{0}\footnote{Corresponding authors (liuji04@baidu.com and jiajuncheng@suda.edu.cn).}, Juncheng Jia$^1$\footnotemark[1], Yang Zhou$^3$, Huaiyu Dai$^4$, Dejing Dou$^2$
\affiliations
$^1$School of Computer Science and Technology, Soochow University, 
$^2$Baidu Inc., China,\\
$^3$Auburn University,
$^4$North Carolina State University, United States
}

\begin{document}

\maketitle

\begin{textblock*}{8cm}(6.5cm,18cm) 
   {\huge To appear in IJCAI}
\end{textblock*}

\begin{abstract}
Despite achieving remarkable performance, Federated Learning (FL) suffers from two critical challenges, i.e., limited computational resources and low training efficiency.
In this paper, we propose a novel FL framework, i.e., \TheName{}, with two original contributions, to exploit the insensitive data on the server and the decentralized data in edge devices to further improve the training efficiency.
First, a dynamic server update algorithm is designed to exploit the insensitive data on the server, in order to dynamically determine the optimal steps of the server update for improving the convergence and accuracy of the global model.
Second, a layer-adaptive model pruning method is developed to perform unique pruning operations adapted to the different dimensions and importance of multiple layers, to achieve a good balance between efficiency and effectiveness.
By integrating the two original techniques together, our proposed FL model, \TheName{}, significantly outperforms baseline approaches in terms of accuracy (up to 4.8\% higher), efficiency (up to 2.8 times faster), and computational cost (up to 61.9\% smaller).
\end{abstract}

\section{Introduction}


While a large amount of data are generally dispersed over numerous edge devices, 
some data may be available at the server side to facilitate more efficient processing 
\cite{zhao2018federated}. 
Due to the concerns of data privacy and security, multiple legal restrictions 
\cite{GDPR,CCPA}
have been put into practice, which makes it complicated to aggregate the distributed sensitive data into a single server or data center. With the size of training data having a significant influence on the quality of machine learning models \cite{li2021practical}, \textbf{Federated Learning} (\textbf{FL}) \cite{mcmahan2017communication,liu2022distributed} becomes a promising approach to collaboratively train a model without transferring raw data. 

Conventional FL was proposed to train a global model using non-Independent and Identically Distributed (non-IID) data on mobile devices \cite{mcmahan2017communication}. During the training process of FL, the raw training data is kept within each device while the gradients or the weights of a model are communicated. FL typically utilizes a parameter server architecture \cite{liu2022distributed}, where a parameter server (server) coordinates the distributed training process. The training process generally consists of multiple rounds, each of which is composed of three steps. First, the server randomly selects several devices and sends the global model to the selected devices. Second, each selected device updates the global model with local data and uploads the updated model to the server. Third, the server aggregates all uploaded local models to generate a new global model. These steps are repeated until the global model converges or predefined conditions are attained.

Although FL helps preserve the privacy and the security of distributed device data, two unsolved challenges limit the applicability of the FL paradigm in real-world scenarios. On the one hand, although a large amount of data are collected from numerous devices, the data over a single device may be limited, which results in low effectiveness of local training. On the other hand, local devices often have limited computing capacity and communication capacity \cite{Li2018Learning}, which leads to low efficiency in the local training process. 

While sensitive data are not allowed to be transferred among the devices and the server, some insensitive data may reside in the servers or the cloud 
\cite{yoshida2020hybrid}
, e.g., Amazon Cloud \cite{Amazon}, which can be exploited to address the low-effectiveness challenges.
Existing works that introduce the insensitive server data to improve the accuracy of the global model generally encounter high communication costs due to data transfer between the devices and the server \cite{zhao2018federated}. 
In addition, they may suffer from low efficiency due to simple processing of the server data as that in an ordinary device \cite{yoshida2020hybrid} or knowledge transfer for heterogeneous models 
\cite{lin2020ensemble}.

Pruning techniques are adopted within the training process of FL to reduce the communication costs between the devices and the server, as well as to accelerate the training process \cite{jiang2019model}. However, most model pruning approaches in the FL context are lossy pruning methods with a single pruning strategy for all layers of neural networks, without considering their individual dimensions and importance \cite{lin2020hrank}.
The aforementioned straightforward model pruning strategies may fail to generate a good approximation to prune appropriate parts and thus lead to non-trivial failure probability of FL training \cite{zhao2018federated}.

In this paper, we propose \textbf{\TheName{}}, an efficient federated learning framework that collaboratively trains a global model using the device data and the server data. To address the two challenges in FL, we take advantage of the server data and the computing power of the server to improve the performance of the global model while considering the non-IID degrees of the server data and the device data. The non-IID degree represents the difference between the distribution of a dataset and that of all the devices. 

Under this framework, we propose an FL algorithm, i.e., \TheAlgoName{}, to dynamically update the model using the device data and the server data. Within \TheAlgoName{}, we propose a method to dynamically adjust the centralized training within the server according to the accuracy of the global model and the non-IID degrees. Furthermore, we design an adaptive model pruning method, i.e., \ThePruneName{}, to reduce the size of the global model, so as to further improve the efficiency and reduce the computational and communication costs of the training process while ensuring comparable accuracy. \ThePruneName{} performs unique pruning operations for each layer of specific dimensions and importance to achieve a good balance between efficiency and effectiveness, based on the non-IID degrees and the server data. To the best of our knowledge, we are among the first to incorporate non-IID degrees in the dynamic server update and the adaptive pruning for FL. We summarize our contributions as follows:
\begin{enumerate}
    \item We propose a dynamic FL algorithm, i.e., \TheAlgoName{}, to take advantage of both server data and device data within FL. The algorithm dynamically adjusts the global model according to the accuracy of the global model and the non-IID degrees with normalized stochastic gradients.
    \item We propose an adaptive pruning method, i.e., \ThePruneName{}, which performs unique pruning operations adapted to the different dimensions and importance of layers based on the non-IID degrees, to further improve the efficiency with an accuracy comparable to the original model.
    \item We carry out extensive experiments to show the advantages of our proposed approach. \TheName{} significantly outperforms state-of-the-art approaches on four typical models and two datasets in terms of accuracy, efficiency, and computational cost.
\end{enumerate}

The rest of this paper is organized as follows. In Section \ref{sec:relatedwork}, we present related work. In Section \ref{sec:method}, we propose our framework, i.e., \TheName{}, including \TheAlgoName{} and \ThePruneName{}. In Section \ref{sec:experiments}, we show the experimental results using two typical models. Finally, Section \ref{sec:conclusion} concludes.

\section{Related Work}
\label{sec:relatedwork}

FL is first proposed to collaboratively train a global model with the non-IID data across mobile devices \cite{mcmahan2017communication}. Some works (
\cite{mcmahan2021advances,liu2022Efficient} and references therein) already exist to increase the performance of global models, while they only consider the device data. Because of incentive mechanisms \cite{yoshida2020hybrid} or the convenience of cloud services, some insensitive data may reside on the server and can be exploited for the training process. While the server data can be assumed to be IID by gathering the data from devices and can be treated as that in an ordinary device with a simple average method \cite{yoshida2020hybrid}, the IID distribution of the server data is not realistic, and the upload of device data to the server may incur privacy or security concerns and excessive communication cost \cite{jeong2018communication}. In the existing literature, the server data is mainly exploited for some specific objectives, e.g., heterogeneous models \cite{he2020group}, label-free data 
\cite{lin2020ensemble}, and one-shot model training \cite{li2021practical}, 
based on knowledge transfer methods, while these methods may suffer from low accuracy \cite{he2020group,li2021practical}, or long training time \cite{lin2020ensemble}. 
The server data could also be transferred to the devices to further improve the model performance 
\cite{zhao2018federated}, 
which incurs significant communication costs. Finally, the server data can be exploited to select relevant clients for the training \cite{nagalapatti2021game}, which is orthogonal to and can be combined with our approach. 

Model pruning can help reduce the computation and communication cost in the training process 
of FL \cite{jiang2019model}, for which the server data is rarely considered. There are two major types of pruning, i.e., weight (unstructured) pruning and filter (structured) pruning. With unstructured pruning, some parameters are set to 0 without changing the model structure. While it can achieve accuracy comparable to the original model \cite{zhang2021validating} and helps reduce communication costs in FL \cite{jiang2019model}, unstructured pruning has limited advantages on general-purpose hardware in terms of computational cost \cite{lin2020hrank}. In contrast, structured pruning directly changes the structure of the model by removing some neurons (filters in the convolution layers). While it can significantly reduce both computational cost and communication cost \cite{lin2020hrank}, it is complicated to determine the number of filters to preserve in the structured pruning method, and the existing methods are not well adapted to FL. Gradient compression or sparsification can be exploited in FL to reduce communication cost \cite{konevcny2016federated}, which has been extensively researched in literature and won't be further explored in this paper.


\section{Method}
\label{sec:method}

In this section, we first present the system model of our framework, i.e., \TheName{}. Then, we present our FL algorithm, i.e., \TheAlgoName{}, for the collaborative FL model update, and our adaptive model pruning method, i.e., \ThePruneName{}, to reduce the computational and communication cost.

\begin{figure}[!htbp]
\centering
\includegraphics[width=1\linewidth]{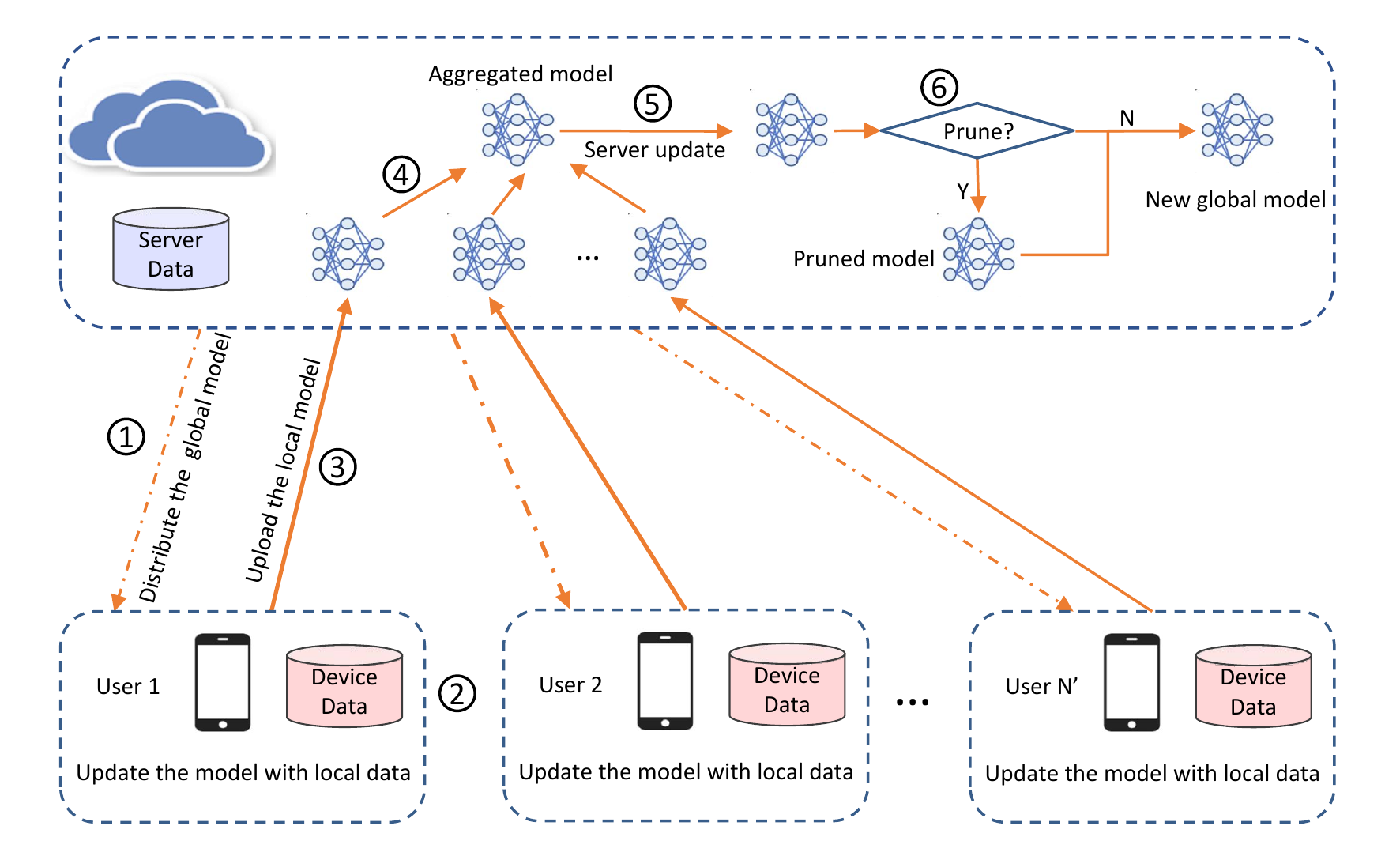}
\vspace{-6mm}
\caption{The training process of \TheName{} Framework.}
\vspace{-6mm}
\label{fig:framework}
\end{figure}

\subsection{System Model}
\label{subsec:systemModel}

As shown in Figure \ref{fig:framework}, we consider an FL system composed of a server and $N$ edge devices.
We assume that the server is much more powerful than the devices. The server data resides in the server, and the device data is distributed in multiple devices, both of which can be locally exploited for the training process of FL without being transferred.
We assume that Device $k$ has a local dataset $D_k = \{x_{k,j}, y_{k, j}\}_{j=1}^{n_k}$, consisting of $n_k$ samples. $D_0$ denotes the server data. $x_{k,j}$ is the $j$-th input data sample on Device $k$, and $y_{k,j}$ is the label of $x_{k,j}$. We denote the input data sample set by $\mathcal{X}$ and the label set by $\mathcal{Y}$.
The objective of the FL training can be formulated as follows:
\vspace{-3mm}
\begin{equation}
\vspace{-2mm}
\min_{w}F(w)\textrm{, with }F(w)\triangleq\frac{1}{n}\sum_{k = 1}^N n_k F_k(w),
\label{eq:problem}
\end{equation}
where $w$ is the parameters of the global model, $F_k(w)\triangleq\frac{1}{n_k}\sum_{\{x_{k,j},y_{k,j}\} \in \mathcal{D}_k} f(w,x_{k,j},y_{k,j})$ is the local loss function of Device $k$, and the loss function $f(w,x_{k,j},y_{k,j})$ captures the error of the model on the data pair $\{x_{k,j},y_{k,j}\}$.

While the data is generally non-IID in FL \cite{mcmahan2021advances}, we exploit the Jensen–Shannon (JS) divergence~\cite{fuglede2004jensen} to represent the non-IID degree of the data on a device and the server as shown in Formula \ref{eq:js-divergence}:
\vspace{-1mm}
\begin{equation}
\vspace{-1mm}
\mathcal{D}(P_k) = \frac{1}{2}\mathcal{D}_{KL}(P_k||P_m) + \frac{1}{2}\mathcal{D}_{KL}(\overline{P}||P_m),
\label{eq:js-divergence}
\end{equation}
where $P_m = \frac{1}{2}(P_k + \overline{P})$, $\overline{P} =\frac{\sum_{k = 1}^{N}n_k P_k}{\sum_{k = 1}^{N}n_k}$, $P_k = \{P_k(y)|y \in \mathcal{Y}\}$ with $P_k(y)$ representing the probability that a data sample corresponds to Label $y$ in Device $k$ or the server ($k=0$), and $\mathcal{D}_{KL}(\cdot||\cdot)$ is the Kullback-Leibler (KL) divergence~\cite{kullback1997information} defined in Formula \ref{eq:kl-divergence}: 
\vspace{-3mm}
\begin{equation}
\vspace{-2mm}
\mathcal{D}_{KL}(P_i||P_j)= \sum_{y \in \mathcal{Y}}P_i(y)\log(\frac{P_i(y)}{P_j(y)}).
\label{eq:kl-divergence}
\end{equation}
A higher non-IID degree indicates that the data distribution of a specific device or the server differs from the global distribution more significantly. While the data cannot be transferred between the server and devices, we assume that the statistical meta information, i.e., $P_k$ and $n_k$, can be shared between the server and devices during the training process, which incurs much less privacy concern~\cite{lai2021oort}.

As shown in Figure \ref{fig:framework}, the training process of \TheName{} consists of multiple rounds, and each round is composed of 6 steps. In Step \textcircled{1}, the server randomly selects a fraction ($\mathscr{D}^t$ with $t$ representing the round) of devices ($\mathscr{D}$) to train a global model and distributes the global model to each device. Then, the model is updated using the local data of each device in Step \textcircled{2}. The updated models are uploaded to the server in Step \textcircled{3}, and aggregated using FedAvg \cite{mcmahan2017communication} in Step \textcircled{4}. Afterward, the model is updated (server update) using the server data in Step \textcircled{5}, which is detailed in Section \ref{subsec:serverUpdate}. Finally, in a specific round, the model is pruned using the statistic information of both the server data and the device data, which is explained in Section \ref{subsec:pruning}. Steps \textcircled{1} - \textcircled{4} are similar to those of conventional FL steps, while we propose using the server data to dynamically update the global model (Step \textcircled{5}) so as to improve the accuracy and an adaptive model pruning method based on both the server data and the device data (Step \textcircled{6}) to accelerate the training of FL.

\subsection{Server Update}
\label{subsec:serverUpdate}

In this section, we propose our FL algorithm, i.e., \TheAlgoName{}, which takes advantage of both server data and device data to update the global model. \TheAlgoName{} exploits the server data to dynamically determine the optimal steps of the server update while considering the non-IID degrees of the server data and the device data in order to improve the convergence and accuracy of the global model. 

While the size of server data may be much bigger than that within a single device, the naive update of the aggregated model with the stochastic gradients generated from the server data may suffer from objective inconsistency \cite{wang2020tackling}. In order to reduce the objective inconsistency, we normalize the stochastic gradients calculated based on the server data, inspired by \cite{wang2020tackling}. Then, the model update in \TheAlgoName{} is defined in Formula \ref{eq:FedGSS}. 
\vspace{-1mm}
\begin{equation}
\vspace{-1mm}
w^t = w^{t - \frac{1}{2}} - \tau_{eff}^{t - 1}\eta \overline{g}_0^{(t - 1)}(w^{t - \frac{1}{2}}),
\label{eq:FedGSS}
\end{equation}
where $w^t$ represents the weights of the global model at Round $t$, $w^{t - \frac{1}{2}}$ represents the weights of the global model after aggregating the models from devices as defined in Formula \ref{eq:FedGSS-device} \cite{mcmahan2017communication}, $\tau_{eff}^{t - 1}$ is the effective step size for the server update defined in Formula \ref{eq:effectiveStep}, $\eta$ is the learning rate, $\overline{g}_0^{t}(\cdot)$ is the normalized stochastic gradient on the server at Round $t$ and defined in Formula \ref{eq:normalization}.
\vspace{-1mm}
\begin{equation}
\vspace{-1mm}
w^{t - \frac{1}{2}} = \sum_{k \in \mathscr{D}^t}\frac{n_k}{n'} (w^{t - 1} - \eta g_k^{t - 1}(w^{t - 1})),
\label{eq:FedGSS-device}
\end{equation}
where $n' = \sum_{k \in \mathscr{D}^t}n_k$ is the total number of samples in the selected devices, $g_k^{t-1}(\cdot)$ is the gradient calculated based on the local data in Device $k$.
\vspace{-2mm}
\begin{equation}
\vspace{-2mm}
\overline{g}_0^{(t-1)}(w^{t - \frac{1}{2}})=\frac{\sum_{i=1}^{\tau}g_0^{(t-1)}(w^{t-\frac{1}{2},i})}{\tau},
\label{eq:normalization}
\end{equation}
where $w^{t-\frac{1}{2},i}$ is the parameters of the updated aggregated model, after $i$ iterations within the batch of the server update at Round $t$, $g_0^{(t-1)}(\cdot)$ represents the corresponding stochastic gradients on the server, and $\tau =\lceil \frac{|n_0| E}{B} \rceil$ is the number of iterations performed within the server update, with $E$ representing the number of local epochs and $B$ representing the batch size. Within each round, the model is updated with multiple iterations by sampling a mini-batch of server data.

\begin{figure}[t]
\vspace{-6mm}
\begin{algorithm}[H]
\caption{Federated Dynamic Server Update (\TheAlgoName{})}
\label{alg:dyamic}
\begin{algorithmic}[1]
\REQUIRE  \quad \\
$\mathscr{D}^t$: The set of selected devices at Round $t$ \\
$\mathcal{D}_k$: The dataset on Device $k$ with 0 representing that on the server \\ 
$w^{t-1}$: The global model at Round $t-1$ \\ 
$E$: The number of local epochs \\
$B$: The local batch size \\
$decay$: The decay rate \\
$P$: The set of data distribution $\{P_k|k \in \{0\} \cup \mathscr{D}^t\}$ with 0 representing the server \\ 
$\eta$: The learning rate
\ENSURE \quad \\
$w^{t}$: The global model at Round $t$ 
\FOR{$k$ in $\mathscr{D}^t$ (in parallel)} \label{line:localUpdateBegin}
\STATE Calculate $g_k^{t - 1}(w^{t - 1})$ using $w^{t-1}$,  $\mathcal{D}_k$
\ENDFOR \label{line:localUpdateEnd}
\STATE Calculate $w^{t - \frac{1}{2}}$ using $w^{t-1}, \eta$ according to Formula \ref{eq:FedGSS-device} \label{line:aggregatedModel}
\STATE Calculate $w^{t}$ using $w^{t-\frac{1}{2}}, decay, E, B, P, \eta$ according to Formula \ref{eq:FedGSS} \label{line:serverUpdate}
\end{algorithmic}
\end{algorithm}
\vspace{-9mm}
\end{figure}

\begin{figure}[t]
\vspace{-6mm}
\begin{algorithm}[H]
\caption{Federated Adaptive Structured Pruning (\ThePruneName{})}
\label{alg:automatic_pruning}
\begin{algorithmic}[1]
\REQUIRE \quad \\
$L$: The list of convolutional layers to prune\\
$\mathscr{D}$: The set of all devices and the server \\
$w$: The initial model\\
$w^*$: The current model at Round $t$\\
$W = [v_1,v_2,\cdots,v_m]$: The list of parameters in the model $w^*$ with $m$ representing the number of parameters
\ENSURE \quad \\
$w'$: The pruned model at Round $t$
\STATE $w' \leftarrow w^*$
\FOR{$k \in \mathscr{D}$ (in parallel)}  \label{line:localPRateBegin} 
        \STATE Calculate the expected pruning rate $p^*_{k}$ \label{line:localPRate}
    \ENDFOR \label{line:localPRateEnd}
\STATE Calculate $p^*$ according to Formula \ref{eq:FedASP} \label{line:fedPRateEnd}
\label{line:eachPRateBegin} 
\STATE $W=[v_{o_1}, v_{o_2}, \cdots, v_{o_R}] \leftarrow$ Sort $W$ in ascending order of $|v|$ \label{line:globalSort}
\STATE $\mathscr{V} = |v_{o_{\lfloor R*p^* \rfloor}}|$ \label{line:threshold}
\FOR{$l \in L$} \label{line:cnnBegin}
    \STATE $W_l = [v_1, v_2,\cdots,v_{q_l}]\leftarrow$ The parameters in Layer $l$ \label{line:localSort}
    \STATE $W'_l \leftarrow [v_1, v_2, \cdots, v_{q'_l}]$ with each $|v_q| < \mathscr{V}$ \label{line:localSelect}
    \STATE $p_l^*\leftarrow \frac{Cardinality(W'_l)}{q_l}$ \label{line:eachPRateEnd} 
    \STATE Calculate the ranks $R_l$ of each filter in Layer $l$ \label{line:rank}
    \STATE Sort the filters according to $R_l$ in an ascending order \label{line:sort}
    \STATE $w'_l \leftarrow$ Preserve the last $d_l - \lfloor p^*_l * d_l \rfloor$ filters in $R_l$ \label{line:prune}
    \STATE $w' \leftarrow$ Replace the $l$-th layer of $w'$ with $w'_l$ \label{line:replace}
\ENDFOR \label{line:cnnEnd}
\end{algorithmic}
\end{algorithm}
\vspace{-9mm}
\end{figure}

As $\tau_{eff}^{t}$ is critical to the training process, we propose to dynamically determine the effective step size based on the accuracy of the aggregated model, the non-IID degrees of the data, and the number of rounds, as defined in Formula \ref{eq:effectiveStep}.
\vspace{-2mm}
\begin{equation}
\vspace{-1mm}
\tau_{eff}^{t} = f'(acc^t) * \frac{n_0 * \mathcal{D}(\overline{P'}^t)}{n_0 * \mathcal{D}(\overline{P'}^t) + n' * \mathcal{D}(P_0)} * \mathcal{C} * decay^t * \tau,
\label{eq:effectiveStep}
\end{equation}
where $acc^t$ is the accuracy (evaluated using the insensitive server data) of the aggregated model at Round $t$, i.e., $w^{t - \frac{1}{2}}$ defined in Formula \ref{eq:FedGSS-device}, $n_0$ is the number of samples in the server data, $\mathcal{D}(\cdot)$ is defined in Formula \ref{eq:js-divergence}, $\overline{P'}^t = \frac{\sum_{k \in \mathscr{D}^t}n_k P_k}{\sum_{k \in \mathscr{D}^t}n_k}$ denotes the distribution of all the data in the selected devices at Round $t$, $P_0$ represents the distribution of the server data, $decay \in (0, 1)$ is used to ensure that the final trained model converges to the solution of Formula \ref{eq:problem}, and $\mathcal{C}$ is a hyper-parameter. $f'(acc)$ is a function based on $acc$.
At the beginning of the training, $acc$ is relatively small, and the value corresponding to $f'(acc)$ is prominent to exploit the server data and the high performance of the server to update the model. At the end of the training, in order to achieve the objective defined in Formula \ref{eq:problem}, $f'(acc)$ becomes small to reduce the influence of the server data. 

\TheAlgoName{} is shown in Algorithm \ref{alg:dyamic}. The local model update is performed in parallel in each selected device (Lines \ref{line:localUpdateBegin} - \ref{line:localUpdateEnd}). Then, the aggregated model is calculated using Formula \ref{eq:FedGSS-device} (Line \ref{line:aggregatedModel}). Afterward, the new global model is updated based on the server data using Formula \ref{eq:FedGSS} (Line \ref{line:serverUpdate}).

\subsection{Adaptive Pruning}
\label{subsec:pruning}

We propose a layer-adaptive pruning method, i.e., \ThePruneName{}, to improve the efficiency of the FL training process. \ThePruneName{} performs unique pruning operations on the server, which are adapted to the different dimensions and importance of multiple layers, based on the non-IID degrees of the server data and the device data. \liu{Please note that the pruning process is carried out at a specific round during the training process.}

\ThePruneName{} is shown in Algorithm \ref{alg:automatic_pruning}. For each device and the server (Lines \ref{line:localPRateBegin} - \ref{line:localPRateEnd}), we calculate the expected pruning rate using the server data and the device data (Line \ref{line:localPRate}). In a device $k$ or the server, given a neural network with initial parameters $W_{k}$, after $T$ rounds of training, we denote the updated parameters by $W'_{k}$ and the difference by $\Delta_{k} = W_{k} - W'_{k}$. Then, we calculate the Hessian matrix of the loss function, i.e., $H(W'_{k})$, and sort its eigenvalues in ascending order, i.e., $\{\lambda^m_{k}|m \in (1, d_{k})\}$ with $d_{k}$ representing the rank of the Hessian matrix and $m$ representing the index of an eigenvalue. We define a base function $B_{k}(\Delta_{k}) = H(W'_{k}) - \triangledown L(\Delta_{k} + W'_{k})$ with $\triangledown L(\cdot)$ representing the gradient of the loss function, and we denote its Lipschitz constant as $\mathscr{L}_{k}$. Inspired by \cite{zhang2021validating}, we find the first $m_{k}$ that satisfies $\lambda_{m_{k+1}} - \lambda_{m_{k}} > 4\mathscr{L}_{k}$ to avoid accuracy reduction, and we calculate the expected pruning rate by $p^*_{k} = \frac{m_{k}}{d_{k}}$, i.e., the ratio between the number of pruned eigenvalues and that of all the eigenvalues. As the expected pruning rate in each device is of much difference because of the non-IID distribution of data, we use Formula \ref{eq:FedASP} to calculate an aggregated expected pruning rate for the entire model on the server (Line \ref{line:fedPRateEnd}).
\vspace{-1mm}
\begin{equation}
\vspace{-1mm}
p^* = \sum_{k = 0}^{n} \frac{ \frac{n_k}{\mathcal{D}(P_k) + \epsilon}}{\sum_{k' = 0}^{n} \frac{n_{k'}}{\mathcal{D}(P_{k'}) + \epsilon}} * p^*_{k},
\label{eq:FedASP}
\end{equation}
where $\epsilon$ is a small value to avoid division by zero. 
We calculate a global threshold value ($\mathscr{V}$), which is used to calculate the pruning rate of each layer. The global threshold value equals the absolute value of the $\lfloor R*p^* \rfloor$-th parameter (Line \ref{line:threshold}) in all the parameters in ascending order (Line \ref{line:globalSort}).
Then, for each convolutional layer (Line \ref{line:cnnBegin}), we calculate the pruning rate of the layer by dividing the number of parameters having inferior absolute values than the threshold value by the total number (Lines \ref{line:localSort} - \ref{line:eachPRateEnd}).
Inspired by \cite{lin2020hrank}, we prune the model based on the ranks of feature maps  (the output of filters) (Lines \ref{line:rank} - \ref{line:replace}).
Let us denote the ranks of feature maps of Layer $l$ by $R_l = \{r_l^j|j \in (1, d_l)\}$, where $d_l$ represents the number of filters in the $l$-th layer. 
While feature maps are almost unchanged for a given model \cite{lin2020hrank}, we assume that the ranks on the server are similar to those in each device. 
Thus, we prune the model with the feature maps on the server.
We calculate (Line \ref{line:rank}) and sort (Line \ref{line:sort}) the ranks of the feature maps in $R_l$. 
We preserve the filters corresponding to the last $d_l - \lfloor p^*_l * d_l \rfloor$ ranks in the sorted $R^l$ (in ascending order), in order to achieve the highest pruning rate $p_l \leq p^*_l$ (Line \ref{line:prune}). Finally, we replace the layer of the original model with the preserved filters (Line \ref{line:replace}). \liu{The process can be applied without the server data by carrying out the execution of Line \ref{line:rank} at a device and transferring the ranks $R_l$ back to the server.}

\begin{figure}[t]
\centering
\subfigure[CNN]{
\includegraphics[width=0.45\linewidth]{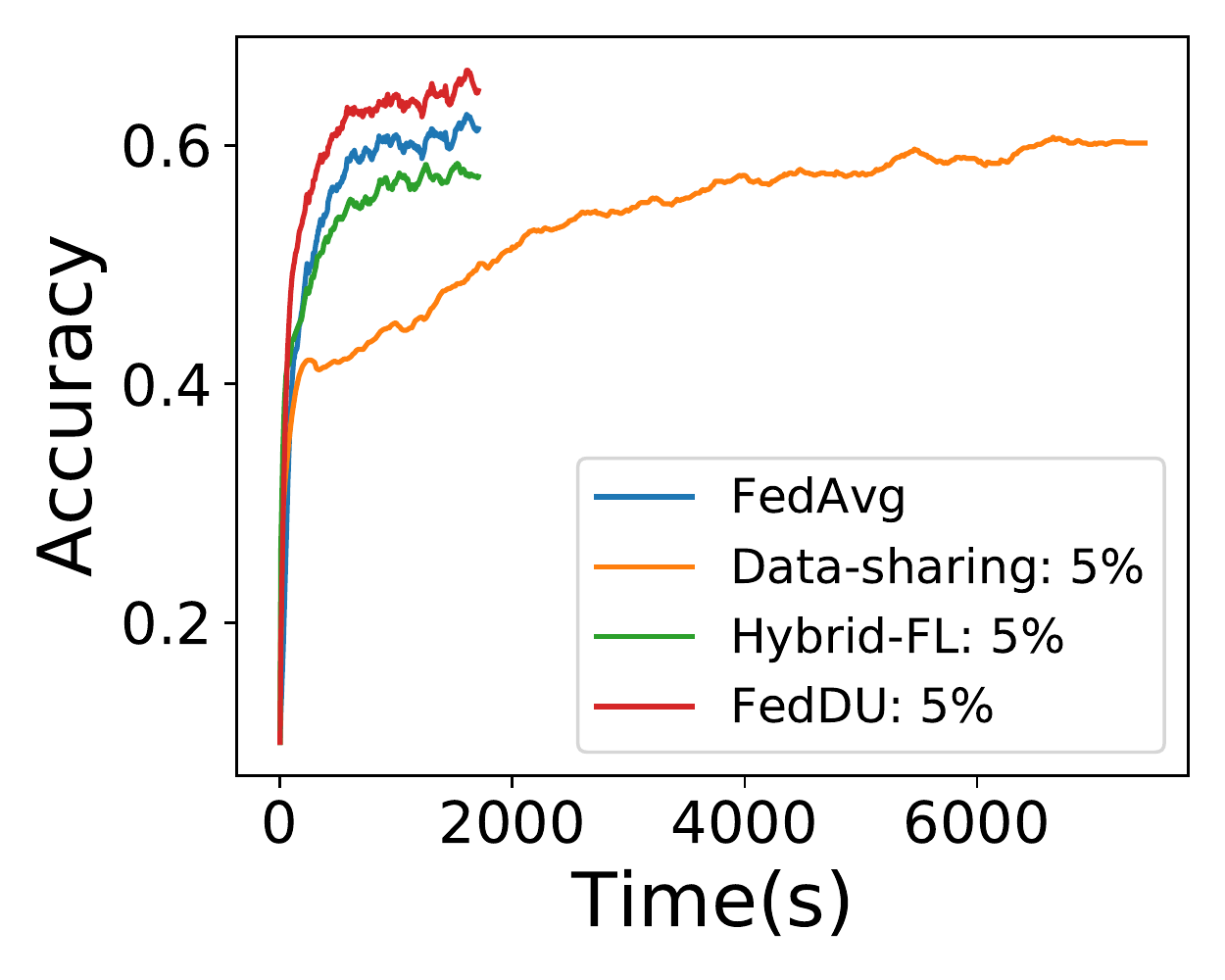}
}
\vspace{-4mm}
\subfigure[CNN]{
\includegraphics[width=0.45\linewidth]{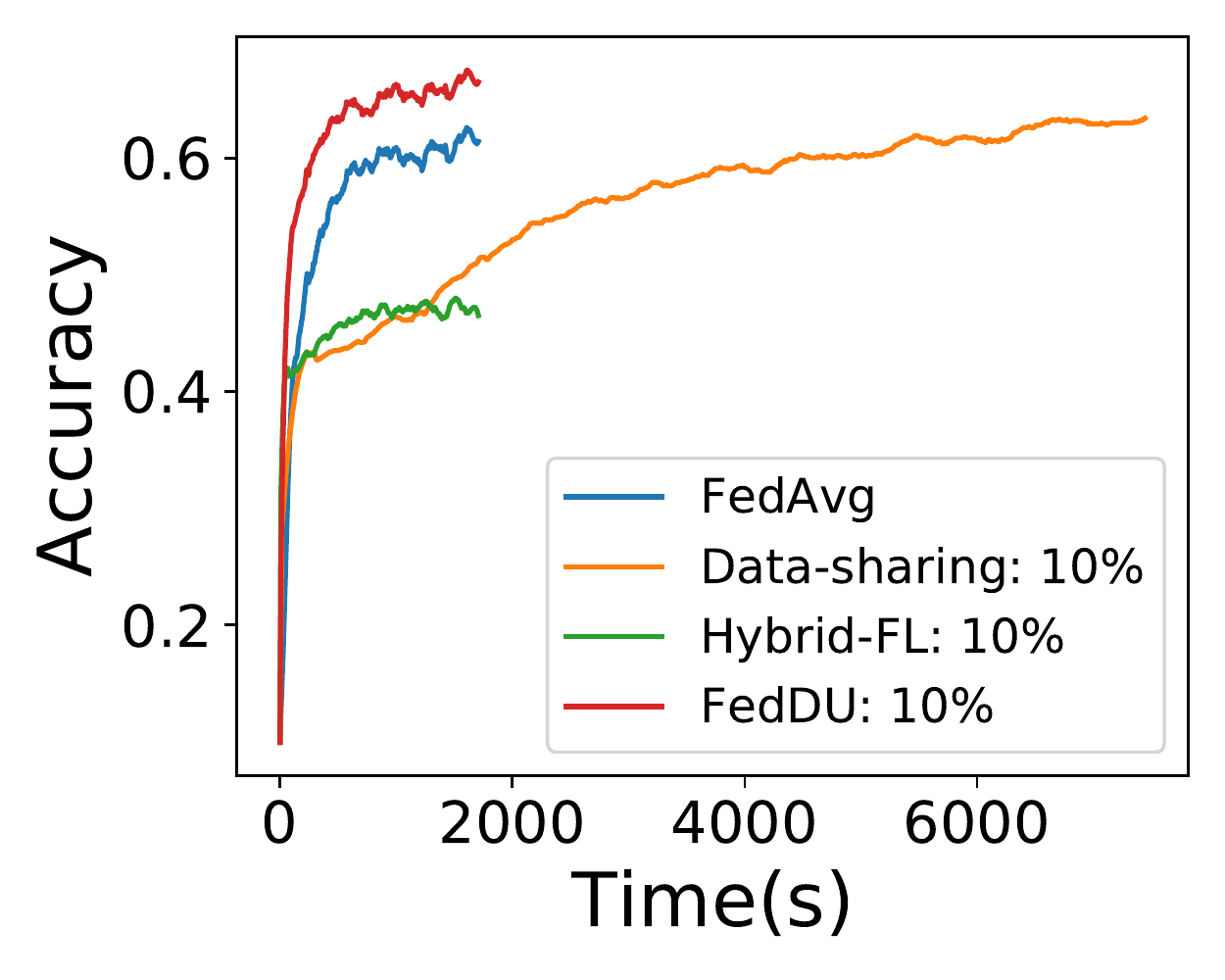}
}

\subfigure[ResNet]{
\includegraphics[width=0.45\linewidth]{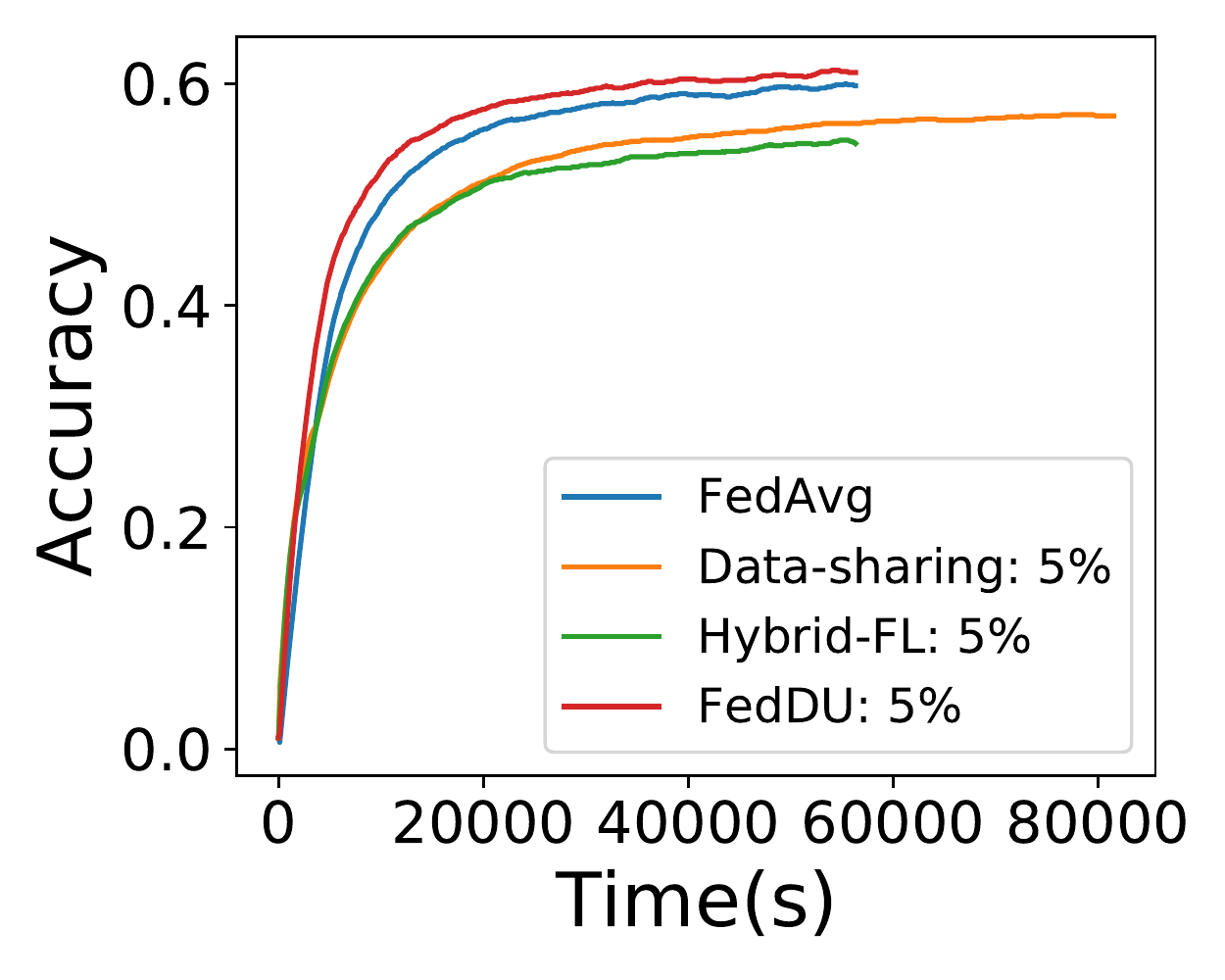}
}
\vspace{-4mm}
\subfigure[ResNet]{
\includegraphics[width=0.45\linewidth]{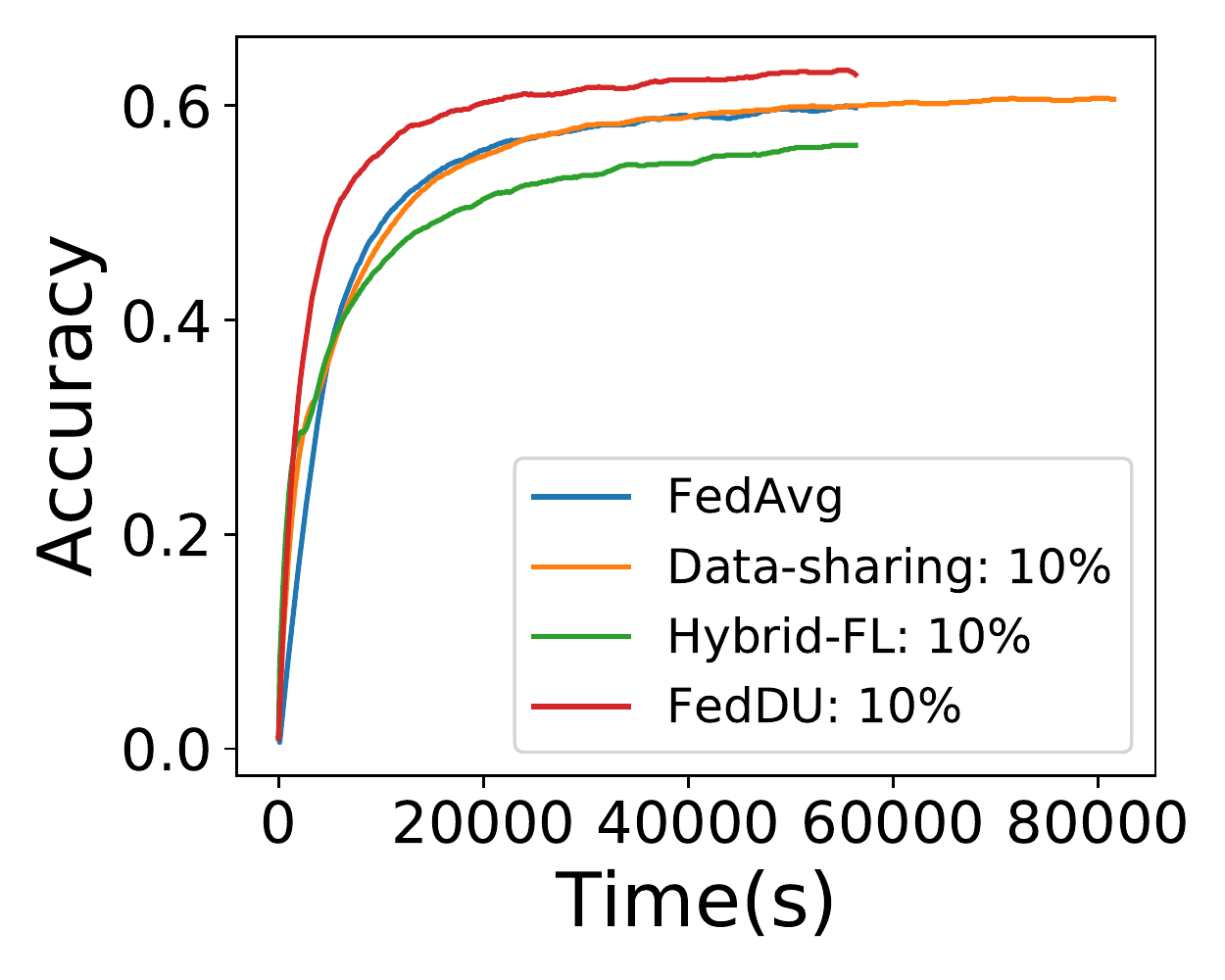}
}

\caption{The accuracy and training time with diverse model update methods corresponding to \TheAlgoName{} with $p = 5\%$ and $p = 10\%$.}
\vspace{-6mm}
\label{fig:cmp_share_l5}
\end{figure}

\section{Experiments}
\label{sec:experiments}

In this section, we present the experimental results to show the advantages of \TheName{} by comparing it with state-of-the-art baselines, i.e., FedAvg~\cite{mcmahan2017communication},  Data-sharing \cite{zhao2018federated},  Hybrid-FL \cite{yoshida2020hybrid}, HRank \cite{lin2020hrank}, IMC~\cite{zhang2021validating}, and PruneFL \cite{jiang2019model}.

\subsection{Experimental Setup}
\label{subsec:expSetup}

We set up an FL system composed of a parameter server and $100$ devices. In each round, we randomly select $10$ devices to participate in the training. We consider the datasets of CIFAR-10 and CIFAR-100 \cite{krizhevsky2009learning} for comparison. We consider four models, i.e., a simple synthetic CNN network (CNN), ResNet18 (ResNet) \cite{he2016deep}, VGG11 (VGG) \cite{simonyan2014very}, and LeNet in our experimentation. 

We report the results of CNN based on CIFAR10 and those of ResNet based on CIFAR100, while the other experimentation results
reveal consistent findings. In Figure \ref{fig:cmp_prunerate_l5}, we report the results of CNN and VGG based on CIFAR10 for the interest of training time. $p$ represents the ratio between the size of data in the server and that in devices. 

\subsection{Evaluation with non-IID Data}

In this section, we first compare \TheAlgoName{} with FedAvg, Data-sharing, and Hybrid-FL in terms of model accuracy. Then, we compare the adaptive pruning method, i.e., \ThePruneName{}, with FedAvg, IMC, HRank, and PruneFL, in terms of model efficiency. Finally, we compare \TheName{}, which contains both \TheAlgoName{} and \ThePruneName{}, with the six state-of-the-art baselines.

\subsubsection{Evaluation on \TheAlgoName{}}

\begin{figure}[t]
\centering
\subfigure[CNN]{
\includegraphics[width=0.45\linewidth]{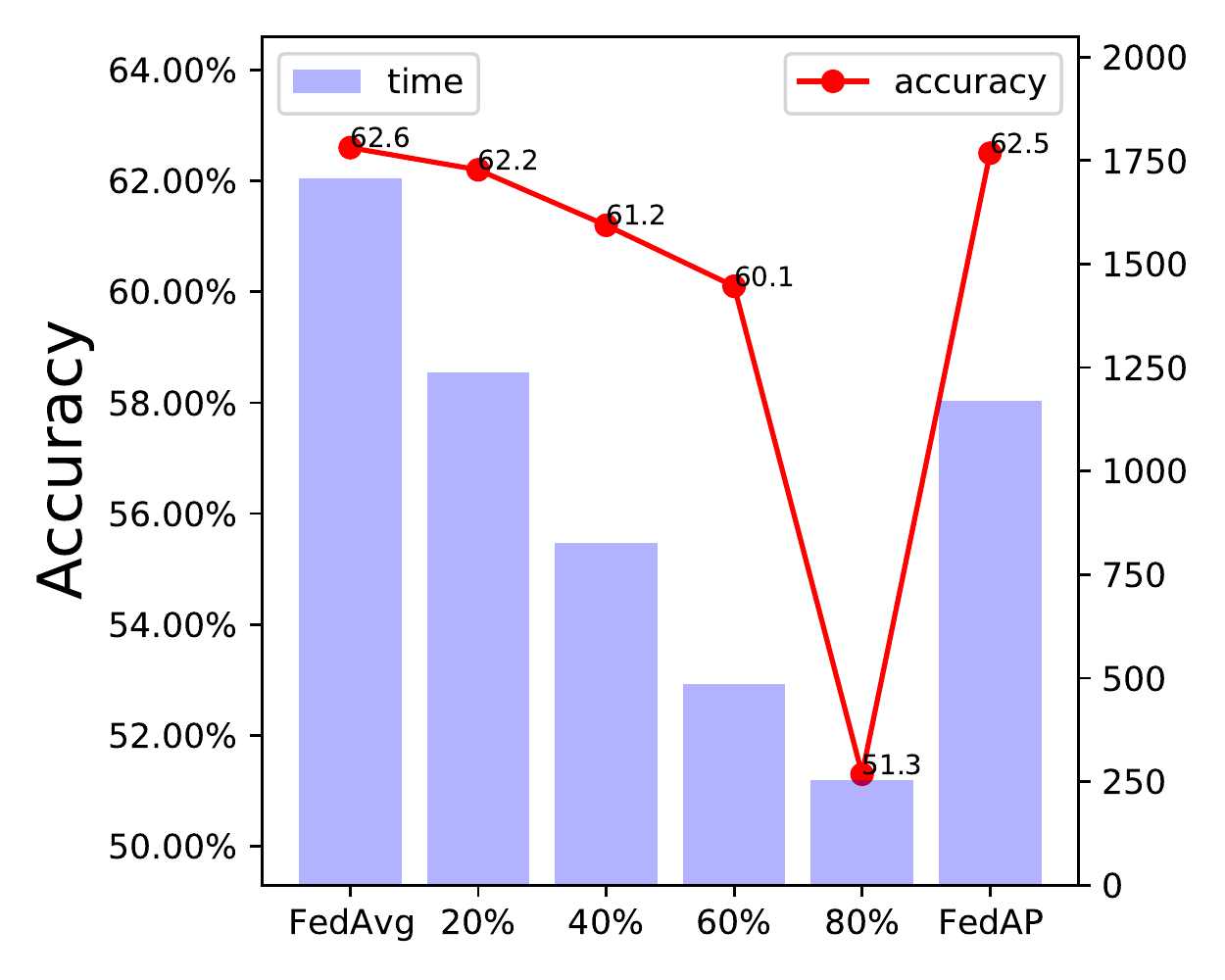}
}
\subfigure[VGG]{
\includegraphics[width=0.45\linewidth]{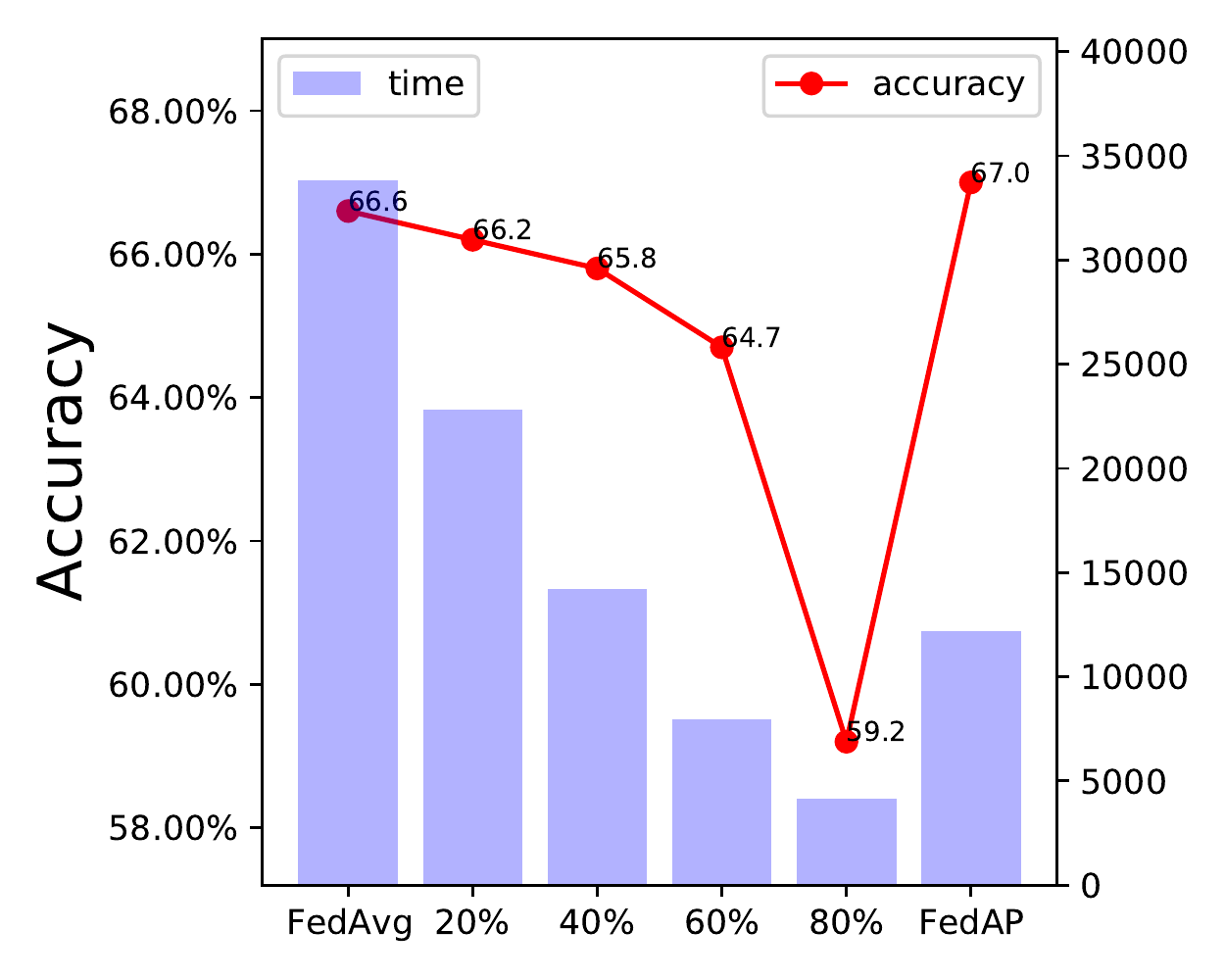}
}
\vspace{-4mm}
\caption{The accuracy and the training time with FedAvg, HRank of diverse pruning rates, and \ThePruneName{}.}
\vspace{-6mm}
\label{fig:cmp_prunerate_l5}
\end{figure}

Data-sharing sends the server data to devices so that devices can mix the local data with the server data to form a dataset with a smaller non-IID degree to increase the accuracy of the global model. Hybrid-FL treats the server as an ordinary device using the FedAvg algorithm.

As shown in Figure \ref{fig:cmp_share_l5}, \TheAlgoName{} leads to a higher accuracy compared with FedAvg (up to 4.9\%), Data-sharing (up to 4.1\%), and Hybrid-FL (up to 19.5\%) for both CNN and ResNet when $p = 5\%$ and $10\%$.  Compared with \TheAlgoName{}, Data-sharing needs to transfer the server data to devices, which may incur a significant communication cost and privacy problems. In addition, Data-sharing needs a much longer training time to achieve the accuracy of 0.6 (up to 15.7 times slower) than \TheAlgoName{} on CNN.

We also evaluate the performance of \TheAlgoName{} with diverse settings of $\tau_{eff}$ and server data.
The dynamic adjustment of $\tau_{eff}$ significantly outperforms a static value (up to 1.7\%) in terms of accuracy. In addition, with various distributions of the server data, we find that when the non-IID degree of the server data is smaller, the global model can get faster performance improvement (up to 6.1 times faster) when comparing the non-IID degree of $0.31$ with that of $9.0 * 10^{-6}$.

\subsubsection{Evaluation on \ThePruneName{}}

In this section, we compare \ThePruneName{} with two adapted methods, i.e., HRank \cite{lin2020hrank}, and IMC \cite{zhang2021validating}, and a state-of-the-art pruning method in FL, i.e., PruneFL \cite{jiang2019model}. We exploit HRank with the server data during the training process of FL with diverse pruning rates, e.g., $0.2$, $0.4$, $0.6$, and $0.8$, of each layer. Similarly, we adapt the IMC method using the server data within the training process of FL. Among the three baseline approaches, HRank is structured, while IMC and PruneFL are unstructured. 
\begin{table}[htbp]
  \centering
  \caption{\textbf{The final accuracy, training time, and computational cost with FedAvg, IMC, PruneFL, and \ThePruneName{}.} ``Accuracy'' represents the accuracy of the final global model. ``Time'' represents the training time (s) to achieve the accuracy of 0.6 for CNN and 0.55 for ResNet. ``Million FLoating-Point Operations (MFLOPs)'' represents the computational cost in devices.}
  \vspace{-3mm}
    \begin{tabular}{|c|c|c|c|c|}
    \hline
    Model & Methods & Accuracy & Time &  MFLOPs \bigstrut\\
    \hline
    \multirow{4}[8]{*}{CNN} & FedAvg & \textbf{0.626} & 838   & 4.6 \bigstrut\\
\cline{2-5}          & IMC   & 0.608 & 1316  & 4.6 \bigstrut\\
\cline{2-5}          & PruneFL & 0.605 & 1536  & 4.6 \bigstrut\\
\cline{2-5}          & \ThePruneName{} & \textbf{0.625} & \textbf{696} & \textbf{3.1} \bigstrut\\
    \hline
    \multirow{4}[8]{*}{ResNet} & FedAvg & 0.600   & 18059 & 557.3 \bigstrut\\
\cline{2-5}          & IMC   & 0.601 & 10141 & 557.3 \bigstrut\\
\cline{2-5}          & PruneFL & 0.560  & 17212 & 557.3 \bigstrut\\
\cline{2-5}          & \ThePruneName{} & \textbf{0.603} & \textbf{5262} & \textbf{246.9} \bigstrut\\
    \hline
    \end{tabular}%
  \label{tab:cmp_nocenter_prune}%
\vspace{-6mm}
\end{table}%
\begin{figure}[t]
\centering
\subfigure[CNN]{
\includegraphics[width=0.45\linewidth]{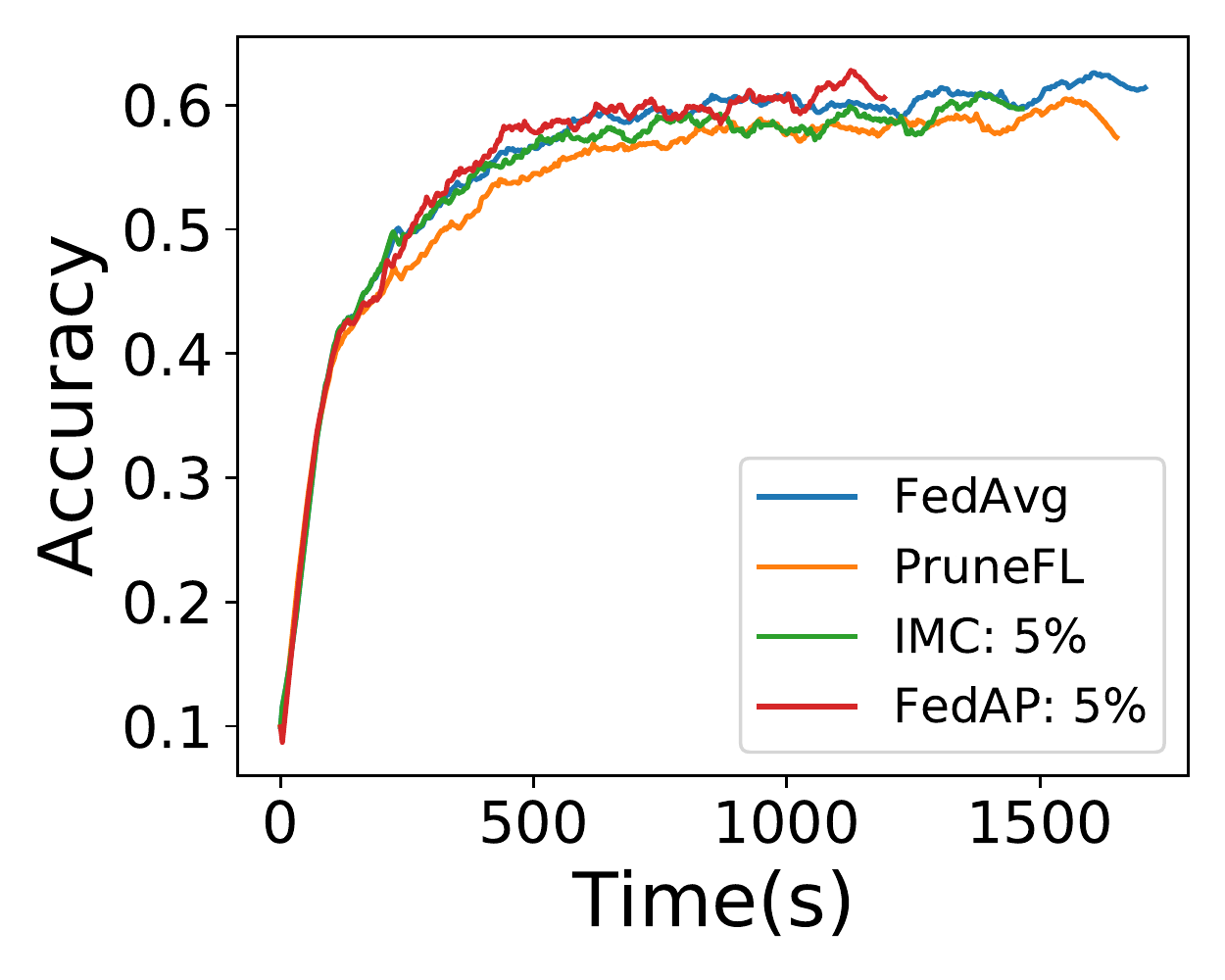}
}
\vspace{-4mm}
\subfigure[CNN]{
\includegraphics[width=0.45\linewidth]{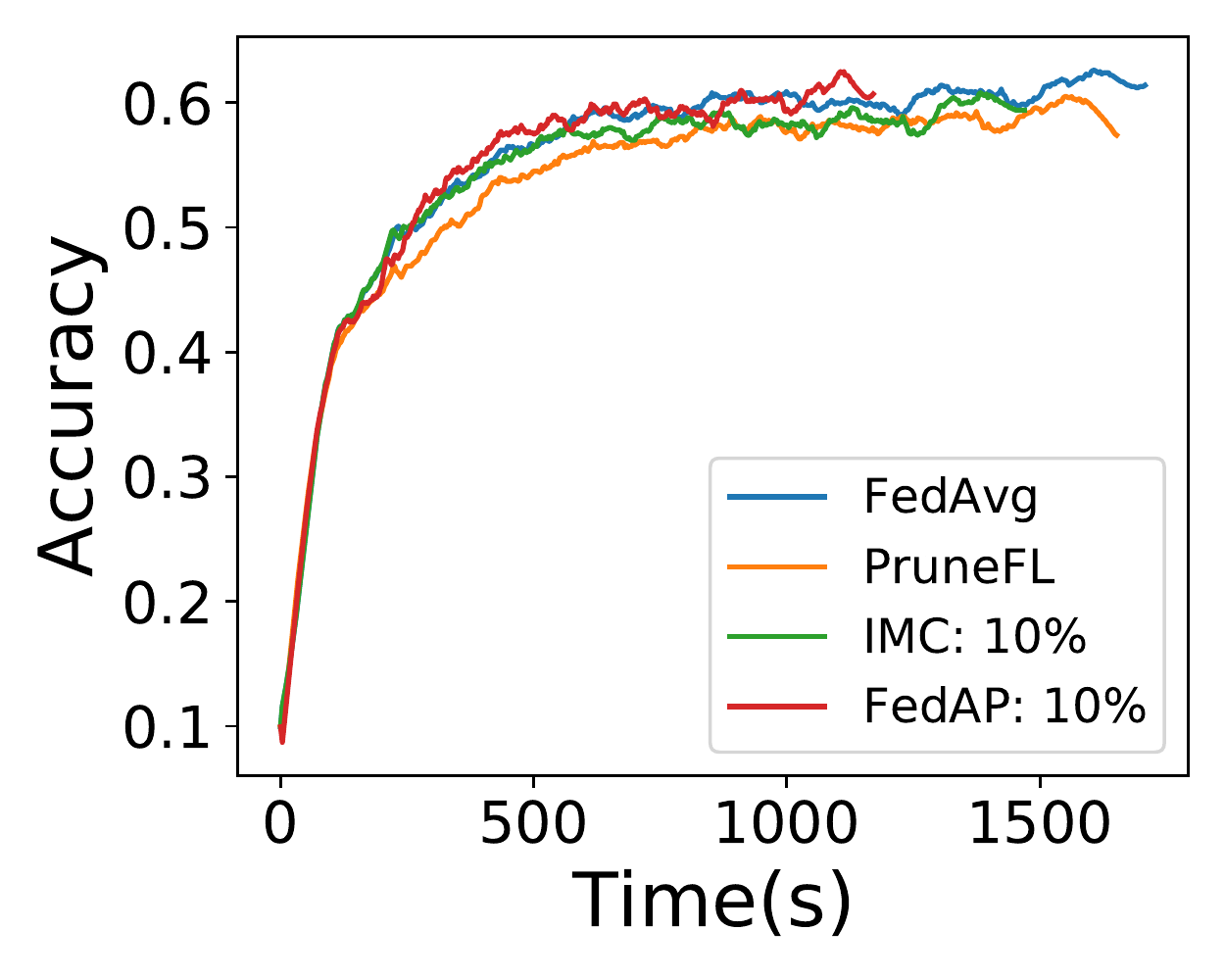}
}
\subfigure[ResNet]{
\includegraphics[width=0.45\linewidth]{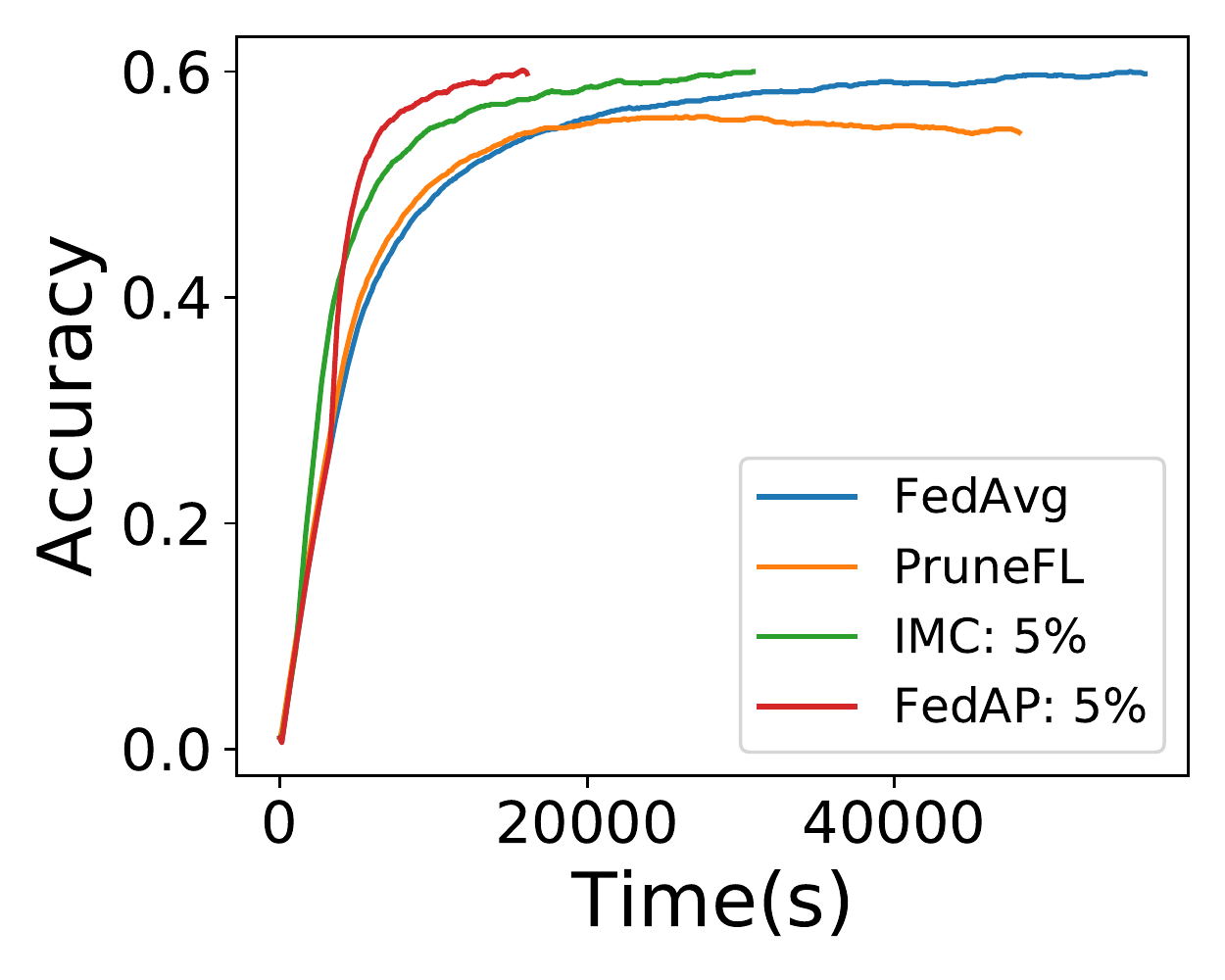}
}
\vspace{-4mm}
\subfigure[ResNet]{
\includegraphics[width=0.45\linewidth]{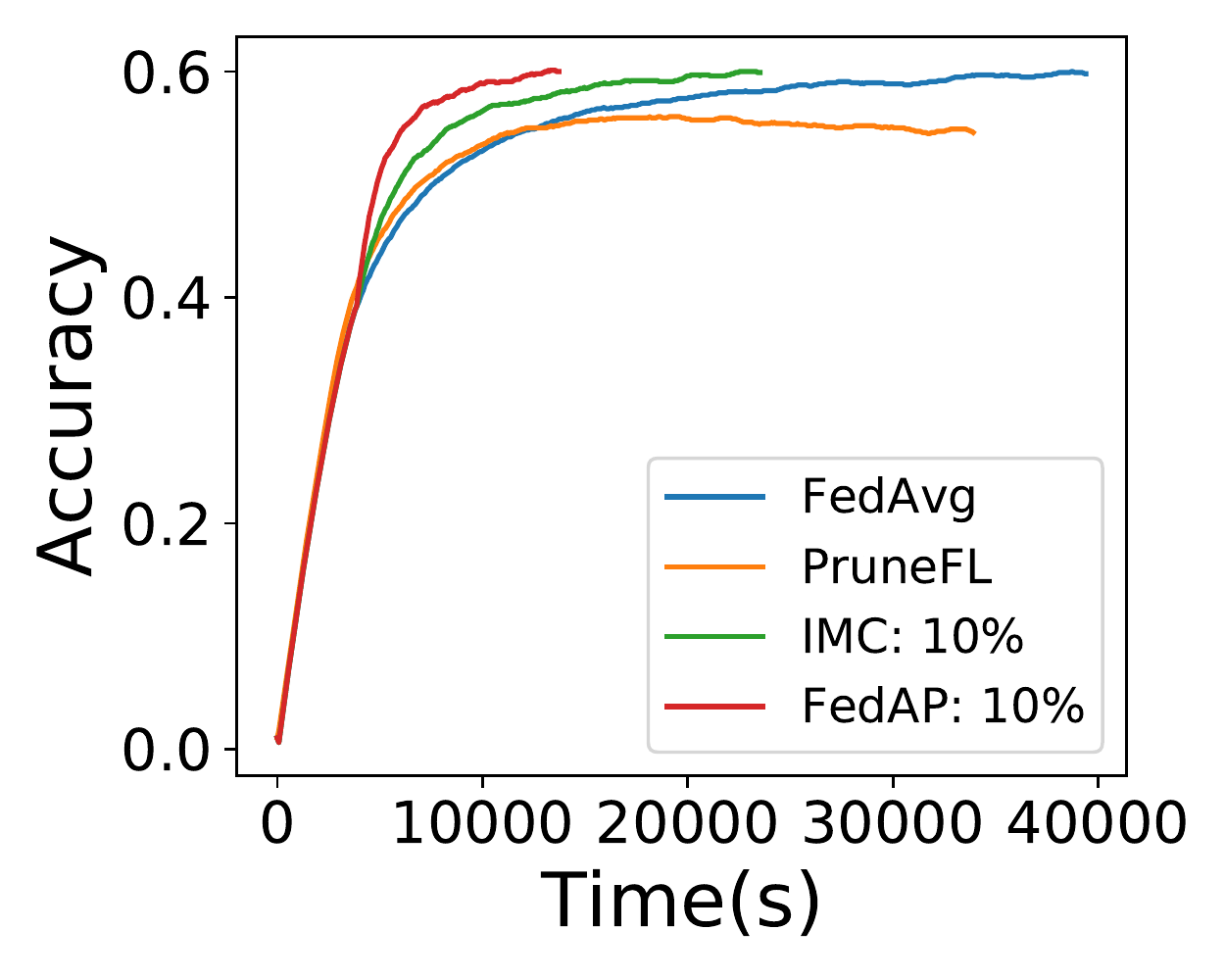}
}
\caption{The accuracy and the training time with FedAvg, IMC, PruneFL, and \ThePruneName{} with $p = 5\%$ and $p = 10\%$. }
\vspace{-8mm}
\label{fig:cmp_imc_prunefl}
\end{figure}
As shown in Figure \ref{fig:cmp_prunerate_l5}, \ThePruneName{} can achieve an appropriate pruning rate  (37.8\% on average for CNN and 72.0\% on average for VGG), which corresponds to a faster training speed (up to 1.8 times faster), while ensuring an excellent accuracy of the model (up to 0.1\% accuracy reduction compared with that of unpruned FedAvg). The training time and the accuracy decrease simultaneously when the pruning rate increases with HRank. It is evident that VGG prefers a pruning rate of 40\%, while CNN favors a pruning rate of 20\% to achieve similar performance as \ThePruneName{}, which reveals that different models correspond to different appropriate pruning rates. Our proposed method, i.e., \ThePruneName{}, can generate adaptive pruning rates for each layer while achieving accuracy comparable with FedAvg. In addition, the accuracy of the pruned model of VGG can be even higher (0.4\%) than the original one, while the training time is one time faster to achieve the accuracy of 0.6. We then compare \ThePruneName{} with FedAvg, IMC, and PruneFL. As shown in Figure \ref{fig:cmp_imc_prunefl}, the training process of \ThePruneName{} is significantly faster than baselines. 
\begin{table}[htbp]
  \centering
  \caption{\textbf{The final accuracy, training time, and computational cost with diverse approaches.} ``Accuracy'' represents the accuracy of the final global model. ``Time'' represents the training time (s) to achieve the accuracy of 0.6 for CNN and 0.55 for ResNet. ``MFLOPs'' represents the computational cost in devices. ``NaN'' represents that the accuracy does not achieve the required accuracy. ``D-S'' represents Data-sharing.}
  \vspace{-3mm}
    \begin{tabular}{|c|c|c|c|c|}
    \hline
    Model & Method & Accuracy & Time & MFLOPs \bigstrut\\
    \hline
    \multirow{6}[12]{*}{CNN} & FedAvg & 0.626 & 838   & 4.6 \bigstrut\\
\cline{2-5}          & D-S   & 0.634 & 4447  & 4.6 \bigstrut\\
\cline{2-5}          & Hybrid-FL & 0.480  & NaN   & 4.6 \bigstrut\\
\cline{2-5}          & IMC   & 0.608 & 1316  & 4.6 \bigstrut\\
\cline{2-5}          & PruneFL & 0.605 & 1536  & 4.6 \bigstrut\\
\cline{2-5}          & \TheName{} & \textbf{0.662} & \textbf{278} & \textbf{2.9} \bigstrut\\
    \hline
    \multirow{6}[12]{*}{ResNet} & FedAvg & 0.600   & 18059 & 557.3 \bigstrut\\
\cline{2-5}          & D-S   & 0.607 & 19114 & 557.3 \bigstrut\\
\cline{2-5}          & Hybrid-FL & 0.563 & 41422 &  557.3 \bigstrut\\
\cline{2-5}          & IMC   & 0.601 & 10141 & 557.3 \bigstrut\\
\cline{2-5}          & PruneFL & 0.560  & 17212 & 557.3 \bigstrut\\
\cline{2-5}          & \TheName{} & \textbf{0.635} & \textbf{4718} & \textbf{250.8} \bigstrut\\
    \hline
    \end{tabular}%
  \label{tab:finalComparison}%
\vspace{-5mm}
\end{table}%
As shown in Table \ref{tab:cmp_nocenter_prune}, \ThePruneName{} achieves the highest accuracy among the pruning methods and corresponds to negligible accuracy reduction (up to 0.1\% compared with unpruned FedAvg), while the training time is significantly shorter to achieve the accuracy of 0.55 (2.4 times faster than FedAvg, up to 92.7\% faster than IMC, and 2.3 times faster than PruneFL). As unstructured pruning methods cannot exploit general-purpose hardware to speed up the calculation, the computational cost remains unchanged. In contrast, \ThePruneName{} leads to a much smaller computational cost (up to 55.7\% reduction).

\subsubsection{Evaluation on \TheName{}}

We compare \TheName{}, consisting of both \TheAlgoName{} and \ThePruneName{}, with the baseline approaches. As shown in Table \ref{tab:finalComparison},
\TheName{} achieves higher accuracy compared with FedAvg (4.8\%), Data-sharing (12.0\%), Hybrid (18.4\%), IMC (6.9\%), and PruneFL (6.3\%). In addition, \TheName{} corresponds to a shorter training time to achieve the target accuracy (up to 2.8 times faster than FedAvg, 15.0 times faster than Data-sharing, 8.8 times faster than Hybrid-FL, 3.7 times faster than IMC, and 4.5 times faster than PruneFL) and a much smaller computational cost (up to 61.9\% reduction for FedAvg, Data-sharing, Hybrid, IMC, and PruneFL).

\section{Conclusion}
\label{sec:conclusion}
In this work, we propose an efficient FL framework based on shared insensitive data on the server, i.e., \TheName{}. \TheName{} consists of a dynamic update FL algorithm, i.e., \TheAlgoName{}, and an adaptive pruning method, i.e., \ThePruneName{}. We exploit both the server data and the device data to train the global model within FL while considering the non-IID degrees of the data. We carry out extensive experimentation with diverse models and real-life datasets. The experimental results demonstrate that \TheName{} significantly outperforms the state-of-the-art baseline approaches in terms of accuracy (up to 4.8\% higher), efficiency (up to 2.8 times faster), and computational cost (up to 61.9\% smaller).

\section*{Acknowledgments}
This work was partially (for J. Jia) supported by the Collaborative Innovation Center of Novel Software Technology and Industrialization.


\bibliographystyle{named}     
\bibliography{bibfile}



\end{document}